\documentclass{article}

\PassOptionsToPackage{numbers, compress}{natbib}

\usepackage[dblblindworkshop, final]{neurips_2026}

\workshoptitle{AI4MetaScience}

\usepackage[utf8]{inputenc} 
\usepackage[T1]{fontenc}    
\usepackage{hyperref}       
\usepackage{url}            
\usepackage{booktabs}       
\usepackage{amsmath}        
\usepackage{amsfonts}       
\usepackage{nicefrac}       
\usepackage{microtype}      
\usepackage{xcolor}         

\usepackage{makecell}    
\usepackage{array}       
\usepackage{tabularx}
\usepackage{graphicx}    
\usepackage{grffile}     
\usepackage{subcaption}

\newcolumntype{P}[1]{>{\raggedright\arraybackslash}p{#1}}

\makeatletter
\if@anonymous
  \newcommand{\thedbfirst}{our \textit{AI in Science Database} (stylized name redacted for double-blind review)}
  \newcommand{\thedb}{our \textit{AI in Science Database}}
  \newcommand{\dbheading}{AI Adoption in Science Database}
  \newcommand{\ourinstitution}{our institution}
  \newcommand{\ourhpc}{an institutional supercomputing cluster}
\else
  \newcommand{\thedbfirst}{the \textit{FutureTech AI in Science Database}}
  \newcommand{\thedb}{the \textit{FutureTech AI in Science Database}}
  \newcommand{\dbheading}{FutureTech AI Adoption in Science Database}
  \newcommand{\ourinstitution}{MIT}
  \newcommand{\ourhpc}{Lincoln Laboratory's Supercloud~\cite{reuther2018interactive}}
\fi

\if@neuripsfinal\else
  \renewcommand{\@noticestring}{Accepted to AI for Meta-Science workshop (NeurIPS 2026).}
\fi
\makeatother

\newcommand{\numfoundationmodels}{2,100}
\newcommand{\nummodelsmatches}{1,629}
\newcommand{\numuniquepapers}{522,248}
\newcommand{\numuniquesentences}{1,432,152}

\newcommand{\pctLLMs}{51\%}
\newcommand{\pctVisionModels}{18\%}
\newcommand{\pctMultimodalModels}{6.9\%}
\newcommand{\pctBiologyModels}{16\%}
\newcommand{\pctImageGenModels}{4.1\%}
\newcommand{\pctSpeechModels}{3.0\%}
\newcommand{\pctGamingModels}{2.7\%}
\newcommand{\pctRoboticsModels}{1.9\%}
\newcommand{\pctVideoModels}{3.3\%}

\newcommand{\cagrLLMs}{41\%}
\newcommand{\cagrVisionModels}{51\%}
\newcommand{\cagrMultimodalModels}{58\%}
\newcommand{\cagrBiologyModels}{152\%}
\newcommand{\cagrImageGenModels}{71\%}
\newcommand{\cagrSpeechModels}{87\%}

\newcommand{\cagrRoboticsModels}{115\%}
\newcommand{\cagrVideoModels}{248\%}

\newcommand{\fmCitingShareLaterYear}{2.5\%}
\newcommand{\fmUsingShareLaterYear}{1.1\%}
\newcommand{\fmCustomizingShareLaterYear}{0.3\%}
\newcommand{\fmAdoptsShareLaterYear}{1.5\%}

\newcommand{\fmCitingShareEarlyYear}{0.19\%}
\newcommand{\fmUsingShareEarlyYear}{0.05\%}
\newcommand{\fmCustomizingShareEarlyYear}{0.004\%}
\newcommand{\fmAdoptsShareEarlyYear}{0.06\%}

\newcommand{\ExtendsEarlyThroughTwentyFourShareCiters}{9.3\%}

\newcommand{\totalAbsoluteUsersTwentyThree}{60,500}
\newcommand{\totalAbsoluteExtendersTwentyThree}{20,700}

\newcommand{\meanBuilderToAdopterFifteen}{5.4x}
\newcommand{\meanAdopterToBuilderTwentyFour}{6.9x}

\newcommand{\adoptVisionModels}{62\%}
\newcommand{\adoptLanguageModels}{25\%}

 \newcommand{\ComputerScienceUseCagr}{19\%}
\newcommand{\ComputerScienceExtendsCagr}{18\%}

\newcommand{\ComputerScienceAdoptsCurrentRate}{19\%}

\newcommand{\ComputerScienceUseExtendUnadjusted}{}

\newcommand{\LinguisticsUseCagr}{26\%}
\newcommand{\LinguisticsExtendsCagr}{40\%}

\newcommand{\LinguisticsAdoptsCurrentRate}{38\%}

\newcommand{\LinguisticsUseExtendUnadjusted}{}

\newcommand{\EngineeringUseCagr}{10\%}
\newcommand{\EngineeringExtendsCagr}{12\%}

\newcommand{\EngineeringAdoptsCurrentRate}{4.3\%}

\newcommand{\EngineeringUseExtendUnadjusted}{}

\newcommand{\MathematicsUseCagr}{20\%}
\newcommand{\MathematicsExtendsCagr}{38\%}

\newcommand{\MathematicsAdoptsCurrentRate}{2.1\%}

\newcommand{\MathematicsUseExtendUnadjusted}{}

\newcommand{\EnvironmentalScienceUseCagr}{14\%}
\newcommand{\EnvironmentalScienceExtendsCagr}{0.6\%}

\newcommand{\EnvironmentalScienceAdoptsCurrentRate}{1.9\%}

\newcommand{\EnvironmentalScienceUseExtendUnadjusted}{}

\newcommand{\MedicineUseCagr}{25\%}
\newcommand{\MedicineExtendsCagr}{15\%}

\newcommand{\MedicineAdoptsCurrentRate}{1.0\%}

\newcommand{\MedicineUseExtendUnadjusted}{}

\newcommand{\PhysicsUseCagr}{21\%}
\newcommand{\PhysicsExtendsCagr}{63\%}

\newcommand{\PhysicsAdoptsCurrentRate}{0.5\%}

\newcommand{\PhysicsUseExtendUnadjusted}{}

\newcommand{\BiologyUseCagr}{107\%}
\newcommand{\BiologyExtendsCagr}{28\%}

\newcommand{\BiologyAdoptsCurrentRate}{1.7\%}

\newcommand{\BiologyUseExtendUnadjusted}{*}

\newcommand{\ChemistryUseCagr}{162\%}
\newcommand{\ChemistryExtendsCagr}{68\%}

\newcommand{\ChemistryAdoptsCurrentRate}{0.9\%}

\newcommand{\ChemistryUseExtendUnadjusted}{*}

\newcommand{\PsychologyUseCagr}{50\%}
\newcommand{\PsychologyExtendsCagr}{55\%}

\newcommand{\PsychologyAdoptsCurrentRate}{1.1\%}

\newcommand{\PsychologyUseExtendUnadjusted}{*}

\newcommand{\MaterialsScienceUseCagr}{29\%}
\newcommand{\MaterialsScienceExtendsCagr}{6.1\%}

\newcommand{\MaterialsScienceAdoptsCurrentRate}{0.6\%}

\newcommand{\MaterialsScienceUseExtendUnadjusted}{}

\newcommand{\EconomicsUseCagr}{25\%}
\newcommand{\EconomicsExtendsCagr}{-30\%}

\newcommand{\EconomicsAdoptsCurrentRate}{0.9\%}

\newcommand{\EconomicsUseExtendUnadjusted}{}

\newcommand{\topThreeFieldsByLatestAdoption}{Linguistics, Computer Science, and Engineering}

\newcommand{\OpenModelPercentage}{48\%}
\newcommand{\ClosedModelPercentage}{8.6\%}
\newcommand{\UnreleasedModelPercentage}{26\%}
\newcommand{\UndocumentedModelPercentage}{18\%}

 \newcommand{\OpenClosedClassificationMinYear}{2019}
\newcommand{\OpenModelExtendsPercent}{40\%}
\newcommand{\ClosedModelExtendsPercent}{18\%}
\newcommand{\OpenModelUsageCAGR}{24\%}
\newcommand{\CloseModelUsageCAGR}{423\%}

\newcommand{\OpenModelExtendsCAGR}{19\%}
\newcommand{\CloseModelExtendsCAGR}{181\%}

\newcommand{\PercentOpenModelAllAdopters}{78\%}

\newcommand{\regressionsYear}{2020}
\newcommand{\tenFoldParameterjournalImpactFactorIncrease}{0.9}
\newcommand{\pValueImpactFactorParameters}{$< 0.1\%$}
\newcommand{\rsquaredImpactFactorParameters}{0.094}
\newcommand{\tenFoldParameterCitationMultiplier}{1.2x}
\newcommand{\pValueLogCitationsParameters}{$< 0.1\%$}
\newcommand{\rsquaredLogCitationsParameters}{0.136}

\newcommand{\SSProportion}{80\%}
\newcommand{\USProportion}{5\%}
\newcommand{\BothSSUSProportion}{15\%}

\newcommand{\impactFactorYear}{2023}

\newcommand{\omittedFields}{Agricultural And Food Sciences, Art, Business, Education, Geography, Geology, History, Law, Philosophy, Political Science, and Sociology}

\newcommand{\LastFiveYrsAdoptionFPOverestimation}{22\%}
\newcommand{\LastFiveYrsExtendsFPOverestimation}{124\%}
\newcommand{\LastFiveYrsCitationsFPCorrectedAdoptionOverestimation}{160\%}

\newcommand{\LinguisticsAverageAgeModelAdopted}{3.3}

\newcommand{\PhysicsAverageAgeModelAdopted}{6.0}

\newcommand{\ChemistryAverageAgeModelAdopted}{3.3}

\newcommand{\MaterialsScienceAverageAgeModelAdopted}{5.8}

\newcommand{\AverageSentencesAdoptPapers}{10}
\newcommand{\AverageAdoptsSentencesAdoptPapers}{2.9}

\newcommand{\adoptionTwentyFourProportionLLMs}{29\%}

\newcommand{\adoptionTwentyFourProportionMultimodalModels}{8.4\%}
\newcommand{\adoptionTwentyFourProportionBiologyModels}{4.5\%}
\newcommand{\adoptionTwentyFourProportionImageGenModels}{5.5\%}
\newcommand{\adoptionTwentyFourProportionSpeechModels}{2.4\%}

\newcommand{\adoptionCAGRLLMs}{35\%}
\newcommand{\adoptionCAGRVisionModels}{8.3\%}
\newcommand{\adoptionCAGRMultimodalModels}{1,155\%}
\newcommand{\adoptionCAGRBiologyModels}{272\%}
\newcommand{\adoptionCAGRImageGenModels}{51\%}
\newcommand{\adoptionCAGRSpeechModels}{43\%}

\newcommand{\overallSizeMultTwentyTwoTwentyFour}{4.1x}
\newcommand{\OnlyLanguageParamGrowthTwentyTwoToFour}{5.1x}

\newcommand{\OnlyMultimodalParamGrowthTwentyTwoToFour}{11x}

\newcommand{\WithoutLanguageMultimodalParamGrowthTwentyTwoToFour}{2.2x}

\title{Science Is Falling Behind the Frontier: Foundation Model Adoption Across Half a Million Papers}

\author{
Ana~Trišović$^{\ast\dagger}$,
Alex~Fogelson$^{\dagger}$,
Janakan~Sivaloganathan$^{}$,
Neil~Thompson$^{\ast}$\\
\small$^{}$MIT FutureTech, 32 Vassar St, Cambridge, MA 02139, USA.\\
\small$^\dagger$These authors share first authorship.\\
\small$^\ast$Corresponding author: ana\_tris@mit.edu, neil\_t@mit.edu
}

\begin{document} 

\maketitle


\begin{abstract}
 We present the first large-scale analysis of AI foundation model \textit{usage} in science -- not just \textit{citations} or \textit{keywords}. We find that adoption has grown rapidly, at nearly-exponential rates, with the highest uptake in \topThreeFieldsByLatestAdoption{}. Vision models are the most used foundation models in science, although language models' share is growing. Open-weight models dominate.  As AI builders increase the parameter counts of their models, scientists have followed suit but at a much slower rate: in 2015, the mean foundation model adopted in science was \meanBuilderToAdopterFifteen{} \textit{larger} than the mean model being built; by 2024 that relationship had reversed, with the mean model built \meanAdopterToBuilderTwentyFour{} larger than the mean model adopted. We also present suggestive evidence that scientists' use of these smaller models may be limiting them from getting the full benefits of AI-enabled science, as papers that use larger models appear in higher-impact journals and accrue more citations.
 \end{abstract}

\noindent At the forefront of the Artificial Intelligence (AI) revolution are foundation models: large pre-trained systems with a broad applicability that contrasts with earlier narrow, task-specific tools~\cite{bommasani_foundation_2023, bommasani2021opportunities}. Although their development demands substantial investments in compute, data, and human capital for training, these models can thereafter be used for a wide range of downstream scientific tasks. 

Foundation models are now proving transformative across diverse scientific domains~\cite{wang_scientific_2023}. In biology and bioinformatics, they are used for processing biological sequence data, enabling advancements in structure prediction, molecular property analysis, drug design, and single-cell omics research~\cite{chen2024foundation, si2024foundation, zimmermann202532}. In physics, they support scientific reasoning and particle event modeling~\cite{barman2025large, birk2024omnijet}. Pre-trained physics-informed models also enhance accuracy for climate modeling and fluid dynamics simulations in health, industry, and climate-related applications~\cite{subramanian2023towards, yu2025physics, 11181506}. Foundation models are also accelerating materials science, using large experimental databases to identify new materials and predict properties such as superconductivity and polymer structures~\cite{yu2025physics, yin2023forge, jablonka2024leveraging}. 

With the widespread release of large language models (LLMs), usage has further expanded to tasks central to everyday scientific practice. LLMs assist with literature review and synthesis, manuscript drafting and editing, code generation and analytical pipeline construction, and even experimental design and interpretation~\cite{gray2024chatgpt, zhang2025exploring, narayan2022can}. Increasingly integrated into routine workflows, LLMs act as versatile research assistants~\cite{binz2025should}. While many uses of foundation models play a supportive role, others such as image segmentation, classification, and predictive modeling directly influence research findings, enabling machine-assisted discovery.

Understanding the prevalence of AI in scientific research is important for evaluating scientists’ productivity~\cite{hao2025ai}, the pace of innovation~\cite{brynjolfsson2017artificial}, and the broader trajectory of economic growth~\cite{besiroglu2024economic}. It also provides insight into how the scientific method is changing with the increasing use of AI, which has broader implications for the validity, transparency, and reproducibility of research findings~\cite{faraboschi_reducing_2024, lawrence_accelerating_2024, gruetzemacher2023leveraging, qiu2023large, yu2025physics}. The significant computational requirements of AI also seem to be deepening existing inequalities in access to research infrastructure~\cite{yin2023forge, gruetzemacher2023leveraging, besiroglu_compute_2024, zenil2023future, yu2025physics, ahmed2023growing,ahmed2020democratization, hao2025role}. Measuring and interpreting trends in foundation model use is therefore essential for informing science policy and funding strategies, as well as for anticipating the long-term economic and societal impacts of AI-driven discovery.

When measuring foundation model usage in science, choosing the wrong indicator can yield a distorted picture. Most previous broad-based studies of AI use citations, keyword-based methods, or abstract classification to identify AI adoption~\cite{duede2024oil, gao2023quantifying, hajkowicz2023artificial, frank_evolution_2019, schmallenbach2024global, hao2025ai, bianchini2025aisupercomputingpoweringwave}. However, using citations as an indicator of adoption conflates meaningful use of a model (e.g., via API or fine-tuning) with papers that merely mention AI for context. Our analysis shows that, over the past five years, citation-based analysis \textit{overestimates} true foundation model adoption by \LastFiveYrsCitationsFPCorrectedAdoptionOverestimation{}. Because text classification approaches, e.g. via LLMs, are imperfect, naively deploying them also creates mismeasurement. We find that using frontier language models without adjustment would lead to a \LastFiveYrsAdoptionFPOverestimation{} overestimation of model adoption and a \LastFiveYrsExtendsFPOverestimation{} overestimation of model customization over that same period. Given these challenges, it is perhaps not surprising that previous work has reported an enormous range of potential adoption rates, ranging from 0.75\% by 2025 to already at 5.3\% by 2021 \cite{hao2025ai,hajkowicz2023artificial}.

Related work has assessed foundation model usage through surveys of scientific practitioners, particularly for LLM usage~\cite{chugunova2025ai, vannoorden2023ai}. That work finds high rates of usage by scientists for various science-related tasks.  While an important measure of adoption in its own right, such surveys do not provide an indication of how central LLMs were to the scientific work. Our analysis, by contrast, provides a strong indicator of importance since ``all elements necessary to allow interpretation and replication of the results" merit inclusion in a paper~\cite{naturemethods2025}. That most scientific usage would fall below this threshold of importance is consistent with other work that shows relatively small impacts of LLMs on non-scientific productivity~\cite{chatterji2025how, humlum2025large}.

To address these limitations, we build an analysis pipeline that directly examines how foundation models are being used in scientific research, and we do extensive data validation to adjust our measurement for the imperfections of language model classifications. Our analysis identifies early adopters, emerging trends, usage disparities, and examines how model domains, scale, and openness shape integration across fields. To support this work, we introduce \thedbfirst, a comprehensive catalog of foundation models and their research adoption patterns.

\section{Tracking Foundation Model Adoption in Science}

\subsection{Foundation Model Definition} \label{sec:fm_definition}

A wide variety of definitions for foundation models have been proposed ~\cite{alfasly2023foundationmodelfoundationmodel, schneider2022foundation}, for example by the Stanford Institute for Human-Centered Artificial Intelligence, the European Union’s AI Act, and the U.S. Department of Homeland Security ~\cite{bommasani2021opportunities, benifei2023proposal, henninger2023foundation}. For our purposes, we focus on a definition that centers on AI's utility as a tool in science:

\begin{quote}
    A foundation model is a pretrained model whose initial development requires substantial investment in data, compute, and expertise, but which can be efficiently adapted and reused across a wide range of downstream tasks. Its value lies not just in its architectural design, but in the broad utility enabled by the upfront training effort.
\end{quote}

\subsection{\dbheading}

To systematically track the use of foundation models in scientific research, we introduce \thedb{} that links foundation models to the scientific publications which cite or apply them.\footnote{Details on this dataset are described in the appendix.}  The database includes the following:

\begin{itemize}
\item \textbf{\nummodelsmatches} \textbf{foundation models} released by 2024, manually curated according to our definition from a snapshot of \numfoundationmodels{} candidate systems. Metadata includes (where available) model origin, openness, number of trainable parameters, and Semantic Scholar identifiers for those models with accompanying publications.

\item \textbf{\numuniquepapers} \textbf{unique papers} citing these foundation models, with associated metadata including authorship, institutional affiliations, and country of origin. Relevant papers and full texts are identified using the Semantic Scholar Academic Graph~\cite{wade2022semantic}.

\item \textbf{\numuniquesentences} \textbf{in-text citations}, including three sentences of context, where foundation models are cited, each annotated with a citation-intent label indicating if and how the model is being applied, and a disambiguation key for cases involving multiple model versions (e.g., different sizes).
\end{itemize}


Together, these components enable rich cross-sectional analyses of foundation model use in science, spanning scientific fields, industry affiliations, geographic regions, and the specific ways in which the models are applied. To characterize the latter, we introduce a custom citation-intention classifier that categorizes references to foundation models into three distinct types:

\begin{itemize}
    \item \textbf{Background Citation}: The model is referenced in a paper to provide context or background, such as in surveys discussing its significance, components, or comparisons with other models. 
    \item \textbf{Model Usage}: The model is applied as a methodological tool without model weight modification, for example via local use or API for content generation, classification, or embeddings. 
    \item \textbf{Model Customization}: The model is used and further customized through fine-tuning or similar methods for domain-specific adaptation.
\end{itemize}

Throughout the text, we refer to \textit{both} usage and customization as collectively representing \textbf{model adoption}. 



\section{Key Trends of Foundation Model Adoption Across Scientific Fields}

\subsection{Foundation model adoption is surging in science}

\begin{figure}[htbp]

    \centering
    \includegraphics[width=\linewidth]{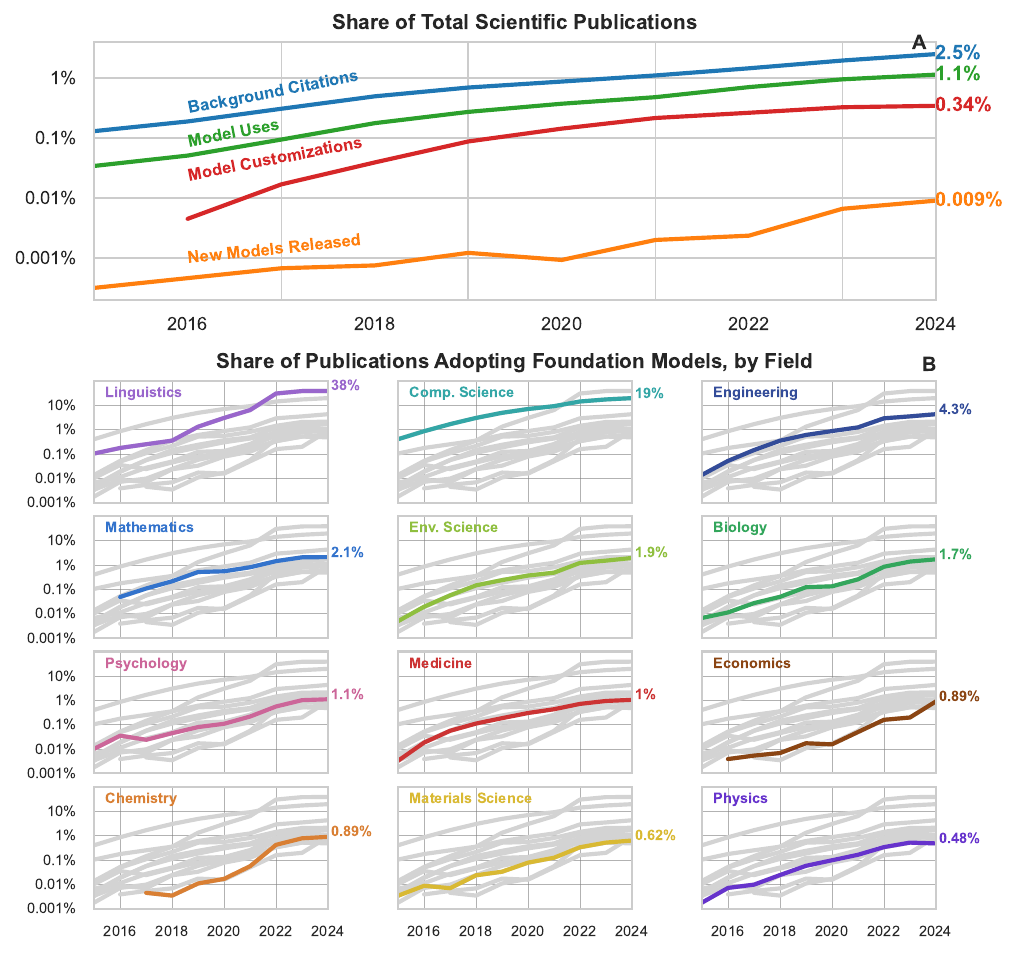}

    \caption{Trends in scientific engagement with foundation models:
    (A) Share of publications citing, using, customizing, or releasing foundation models. Publications with multiple types are categorized by their most technically demanding category (release, then customization, then use, then citation); 
    (B) Share of publications adopting foundation models by field, where adoption is defined as use or customization. Grey lines represent the other academic fields.} \label{fig:overall}
\end{figure}

Since 2016, the use and customization of foundation models has grown rapidly, at near-exponential rates (Figure~\ref{fig:overall}A). During this period, overall adoption grew from \fmAdoptsShareEarlyYear{} of publications to \fmAdoptsShareLaterYear{}. This includes both the usage of foundation models as-is, which grew from \fmUsingShareEarlyYear{} of publications to \fmUsingShareLaterYear{} in 2024, and also the more technically-demanding customization of models, which rose from \fmCustomizingShareEarlyYear{} of publications to \fmCustomizingShareLaterYear{}. Even these modest percentages translate to an enormous number of papers annually -- we estimate that in 2023, \totalAbsoluteUsersTwentyThree{} papers used a foundation model and \totalAbsoluteExtendersTwentyThree{} papers customized them. Such prolific use reveals that foundation models have already become an important tool in modern science.\footnote{Unsurprisingly, the mentions of AI foundation models as background context in science have also grown from \fmCitingShareEarlyYear{} to \fmCitingShareLaterYear{} share of publications.}

As of 2024, the scientific fields with the highest adoption rates were Linguistics (\LinguisticsAdoptsCurrentRate{}), followed by Computer Science (\ComputerScienceAdoptsCurrentRate{}) and Engineering (\EngineeringAdoptsCurrentRate{}). AI adoption has become pervasive, with 17 of the 23 scientific fields we observe reaching 1\% adoption by 2024 (Figure~\ref{fig:overall_fp}). Among fields with sufficiently rich data to analyze, Chemistry and Biology usage grew the fastest in recent years, with a three-year compound annual growth rate (CAGR) of \ChemistryUseCagr{} and \BiologyUseCagr{} for usage of AI models, and \ChemistryExtendsCagr{} and \BiologyExtendsCagr{} rate for customization, respectively (Table~\ref{tab:fields}).\footnote{See Appendix for field selection details.}   Except for Psychology (\PsychologyUseCagr{}), no other field's usage growth exceeded that of Materials Science (\MaterialsScienceUseCagr{}).
Physics recorded the highest growth in model customization among fields with false-positive-adjusted data (\PhysicsExtendsCagr{}), with Chemistry's unadjusted rate higher still (\ChemistryExtendsCagr{}). At the other extreme, customization was essentially flat in Environmental Science (\EnvironmentalScienceExtendsCagr{}) and declined in Economics (\EconomicsExtendsCagr{}). In general, however, many scientists found it valuable to adapt models with their own training data, with \ExtendsEarlyThroughTwentyFourShareCiters{} of all papers citing foundation models from 2016 to 2024 customizing models.

In all scientific fields, researchers tend to use older, more established models that predate their publication by 3 or more years, and in several fields the typical lag is 5 or 6 years. Linguistics (\LinguisticsAverageAgeModelAdopted{} years old on average) and Chemistry (\ChemistryAverageAgeModelAdopted{}) work closest to the frontier, while Physics (\PhysicsAverageAgeModelAdopted{}) and Materials Science (\MaterialsScienceAverageAgeModelAdopted{}) lag furthest behind. Computer Science adopts the highest number of unique models, whereas Economics adopts the fewest, though from a considerably smaller sample (Table~\ref{tab:fields}).



\subsection{Smaller models dominate adoption, large-scale models dominate development}

\begin{figure}[htbp]
  \centering
    \includegraphics[width=\linewidth]{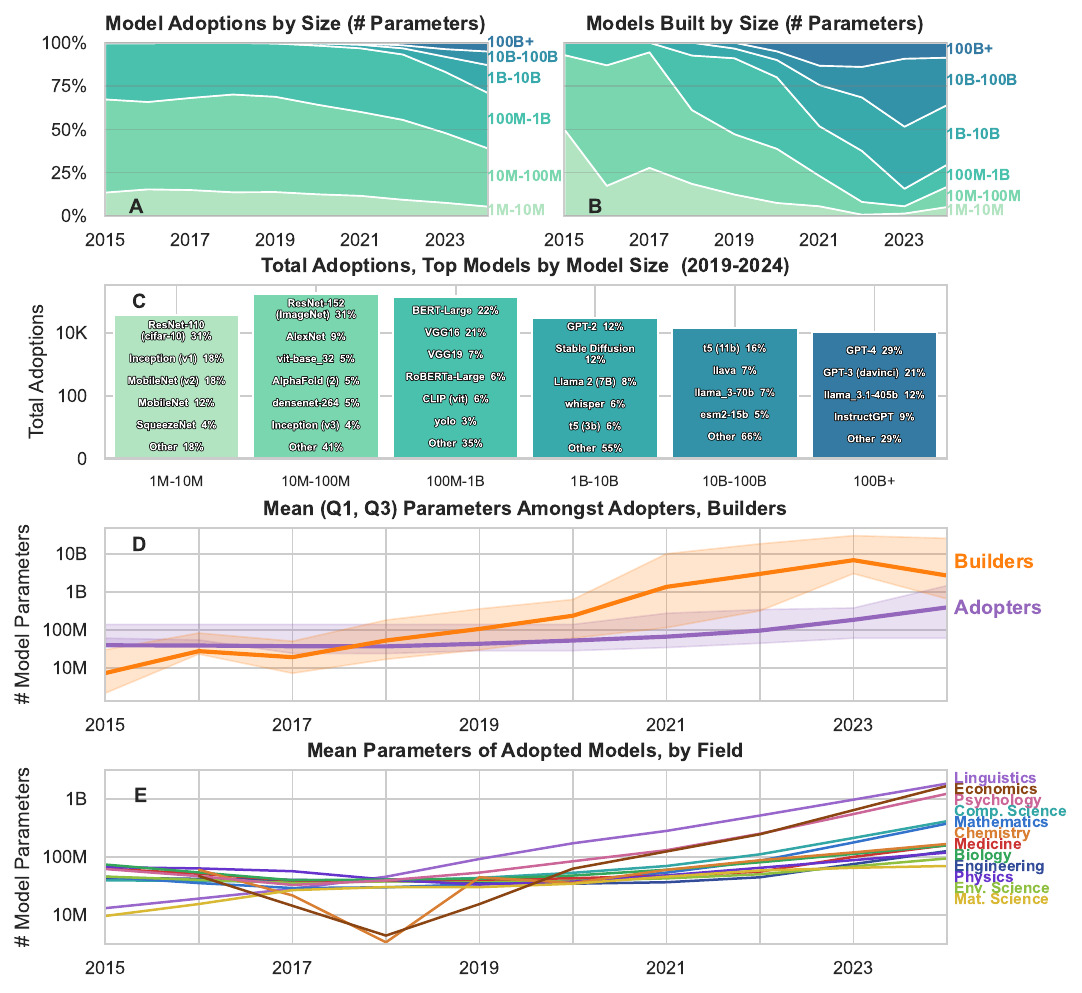}

    \caption{Foundation model adoption across scientific fields: (A) Share of adoptions by parameter count; (B) Share of models being built by parameter count; (C) Total adoption by model size (\# parameters). Most-adopted models are listed in the bar; (D) Trends in mean model size for models built and adopted, with 25th-75th percentiles shaded; (E) Average model size adopted per year, by scientific field.} \label{fig:disciplines}
\end{figure}

Despite the enormous growth in the size of cutting-edge models~\cite{sevilla2022compute, villalobos2022machine}, scientists largely have stuck with small models. Even in 2024, roughly two-fifths of the models adopted had fewer than 100 million parameters and nearly three-quarters had fewer than 1 billion parameters (Figure~\ref{fig:disciplines}A). By contrast, in that same year, nearly 75\% of the foundation models being built had 1 billion or more parameters (Figure~\ref{fig:disciplines}B). 

For many years, the size of models being adopted by scientists increased only gradually. Growth accelerated after 2022, with average model sizes increasing \overallSizeMultTwentyTwoTwentyFour{} by 2024. As might be expected, language and multimodal models played a role in this growth (growing \OnlyLanguageParamGrowthTwentyTwoToFour{} and \OnlyMultimodalParamGrowthTwentyTwoToFour{}, respectively). But other model types also grew significantly over this period, by \WithoutLanguageMultimodalParamGrowthTwentyTwoToFour{}. Despite such growth, the adoption of the largest foundation models in science is limited. We estimate that by 2024 there had been fewer than 10,000 adoptions of 100+ billion parameter models (e.g., GPT4) across all of science (Figure~\ref{fig:disciplines}C).

Up until 2018, adopters typically used larger models than builders, a trend largely driven by widespread use of VGG, ResNet and AlexNet models, which were large models for the time~\cite{simonyan2014very}. Since 2018, scientists have continued the use of similar-sized models even as model builders built ever-larger systems. This has led to a widening gap between the size of foundation models being built and those being used (Figure~\ref{fig:disciplines}D). Over this period the gap has not merely widened but inverted: in 2015 the mean model adopted in science was \meanBuilderToAdopterFifteen{} larger than the mean model being built, whereas by 2024 the mean model built was \meanAdopterToBuilderTwentyFour{} larger than the mean model adopted. Since model capabilities grow with greater model size~\cite{rosenfeld2019constructive, kaplan2020scaling, wei2022emergent}, this suggests that scientists are only partially reaping the benefits of the recent AI advances. A plausible explanation for this limited uptake of cutting-edge models is the cost and technical demands of working with larger systems, a concern frequently raised in policy discussions~\cite{ahmed2023growing, NAIRR2023}. Despite this broad trend, there is evidence that some scientific fields, particularly Linguistics, Economics, and Psychology, have adopted larger models more quickly, suggesting that modern LLMs have particular incremental value for these fields (Figure~\ref{fig:disciplines}E).

\begin{figure}[tbp]
    \centering
    \includegraphics[width=\linewidth]{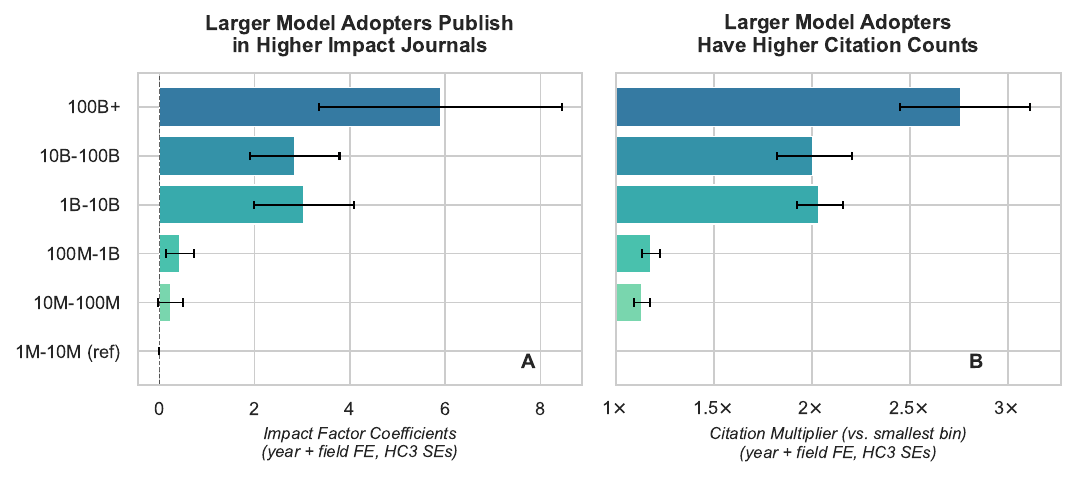}
    \caption{The scholarly impacts of foundation model adoptions: (A) journal impact factor by adopted model size; (B) median citation counts by adopted model size.}
    \label{fig:scholar}
\end{figure}

\subsection{Large model adopters have greater impact}

There is suggestive evidence that the use of large models may lead to stronger scientific outcomes. We observe this in two ways: the usage of larger models is correlated with publication in higher quality journals (Figure~\ref{fig:scholar}A) and with accruing greater numbers of citations (Figure~\ref{fig:scholar}B). These associations hold \textit{after} controlling for publication year and field with regression fixed effects, indicating that they are not simply driven by recency bias. On average since \regressionsYear{}, a ten-fold increase in model parameters corresponds to an article being published in a journal with a \impactFactorYear{} impact factor that is  \tenFoldParameterjournalImpactFactorIncrease{} higher ($p$ \pValueImpactFactorParameters{}, $r^2 =$ \rsquaredImpactFactorParameters{}). It also corresponds to accruing \tenFoldParameterCitationMultiplier{} more citations ($p$ \pValueLogCitationsParameters{}, $r^2 =$ \rsquaredLogCitationsParameters{}).

\subsection{Vision models and open models play a critical role in research}

\begin{figure}[htbp]
  \centering
    \includegraphics[width=\linewidth]{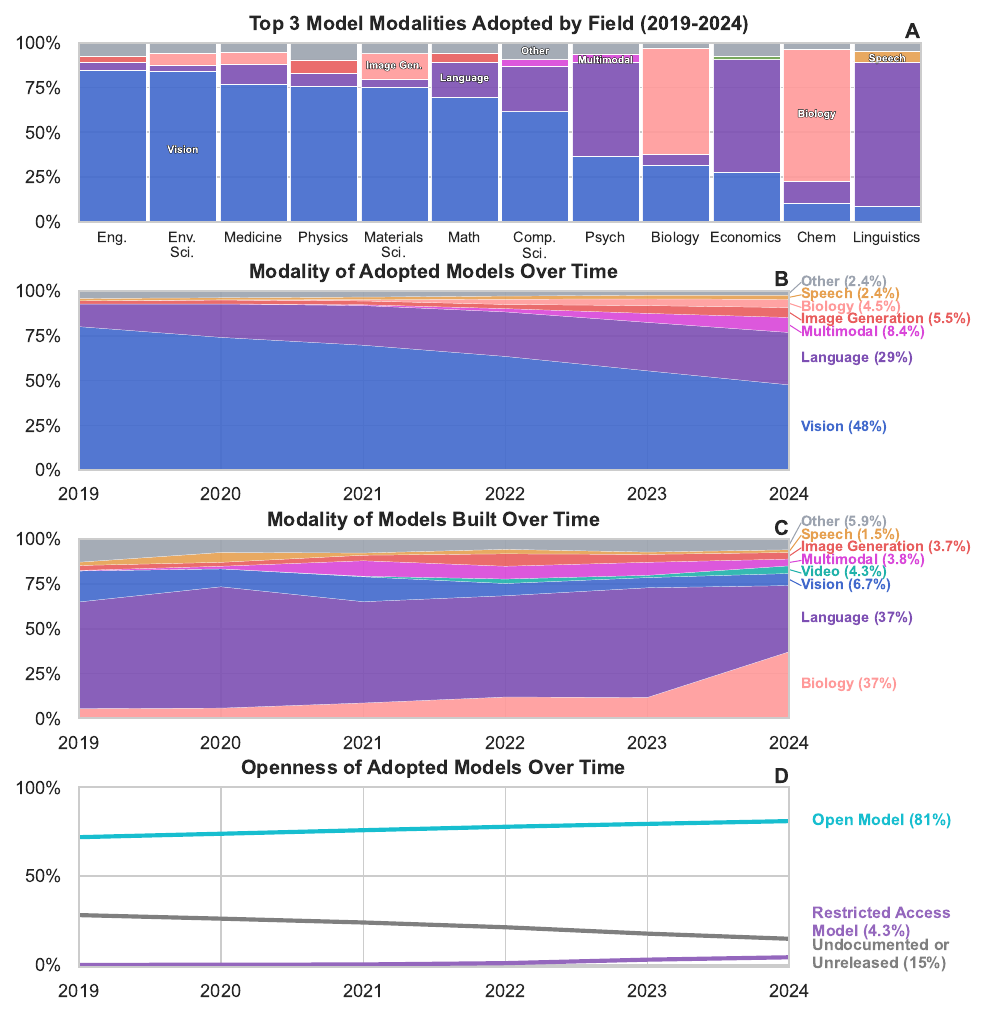}
    \caption{Share of foundation models by (A) adopted modality by field, (B) adopted modality over time, (C) built modality over time, and (D) openness of adopted models over time. In A-C, multiclass models are given equal weight per modality.} \label{fig:modality}
\end{figure}

Looking across scientific fields reveals enormous variation in the types of foundation models being used (Figure \ref{fig:modality}A). In many fields, the dominant models being used are vision models, likely because they allow the larger-scale analysis of image data (for example, as discussed by \cite{zhang2024challenges}). Despite the omnipresence of LLMs in the media, language models are only the dominant foundation models in three fields: Psychology, Economics and Linguistics. Since so much of the data in these fields is textual, this suggests that language model adoption may also be driven by the ability to analyze larger corpora than was previously possible. In Biology and Chemistry, the use of Biology-specific foundation models is dominant.

While the importance of vision models is decreasing over time compared to other types, it still represents \adoptVisionModels{} of all usage overall, and nearly half of all usage in 2024 (Figure~\ref{fig:modality}B). 
Language models are the next most-used foundation models, representing \adoptLanguageModels{} of adoption, with their share growing to \adoptionTwentyFourProportionLLMs{} of usage by 2024. Both of these percentages are part of the overall growth in adoption, with vision model usage growing at an annual rate of \adoptionCAGRVisionModels{}, while language models grow at \adoptionCAGRLLMs{}. Use of other modalities has also grown: multimodal (\adoptionTwentyFourProportionMultimodalModels{}, \adoptionCAGRMultimodalModels{} growth rate), biology (\adoptionTwentyFourProportionBiologyModels{}, \adoptionCAGRBiologyModels{}), image generation (\adoptionTwentyFourProportionImageGenModels{}, \adoptionCAGRImageGenModels{}), and speech (\adoptionTwentyFourProportionSpeechModels{}, \adoptionCAGRSpeechModels{}).

This pattern of scientific adoption represents an entirely different mix of models than builders are creating. Of all the \nummodelsmatches{} foundation models being built, most are language models (\pctLLMs{} overall), followed by vision (\pctVisionModels{}), biology (\pctBiologyModels{}), multimodal (\pctMultimodalModels{}), image generation (\pctImageGenModels{}), video (\pctVideoModels{}), speech (\pctSpeechModels{}), gaming (\pctGamingModels{}), and robotics models (\pctRoboticsModels{}). Figure~\ref{fig:modality}C depicts these proportions over time.\footnote{These categories overlap, as models may span multiple modalities.}\footnote{Additional modality-specific growth rates are provided in Table~\ref{supp:model_cagrs}.}



The adoption of foundation models in science is overwhelmingly of open-weight models, where anyone can download the models and apply or customize them. Across fields since \OpenClosedClassificationMinYear{}, \PercentOpenModelAllAdopters{} of adopted models are open. The ability to freely modify the model also leads to many more customizations of open models than closed ones (\OpenModelExtendsPercent{} of adoptions for open models versus \ClosedModelExtendsPercent{} for closed models).
Despite closed-models limited prevalence, their adoption by scientists is growing more rapidly than open models. From 2020 through 2023, the usage of closed models grew \CloseModelUsageCAGR{} annually (vs \OpenModelUsageCAGR{} for open models). In that same period, customization grew by \CloseModelExtendsCAGR{} (vs. \OpenModelExtendsCAGR{} for open models). Much of this growth in restricted models is of extremely large (100B+ parameter) language models. This growing reliance on restricted-access models raises concerns about the long-term openness and accessibility of AI for scientific research, especially given the well-documented importance of open foundation models~\cite{bommasani_foundation_2023, kapoor_societal_2024, ahmed2023growing}.


This pattern of adoption contrasts strongly with the mix of foundation models being built, where open-weight models comprise \OpenModelPercentage{} of the total, while restricted-access models account for \ClosedModelPercentage{}. There are also \UnreleasedModelPercentage{} of models that are created but not released, and \UndocumentedModelPercentage{} of models for which we could not find documentation to determine openness. 

\subsection{Limitations}

Our findings are limited both by the underlying data that we analyze and the limits of interpreting usage via the text around citations.  For our foundation model data, we build on a set first gathered by Epoch AI~\cite{EpochNotableModels2024}. While a valuable record of foundation models, it likely omits some models as a consequence of fields evolving rapidly. For the identification of citations and adoption, we rely on the Semantic Scholar Academic Graph~\cite{wade2022semantic}, which implies some limitations. Semantic Scholar primarily covers English-language publications, so our analysis largely represents English-language science. It is also updated only periodically and with a lag, hence at the time of writing the total publication count in 2024 is below that of 2023. For this reason, we largely report shares of publications and only report adoption CAGRs as of the end of 2023. Semantic Scholar's categorization of scientific fields is non-exclusive, meaning that individual publications can be present in multiple fields. For our analysis, such papers are counted in each field in which they are categorized.

Methodologically, our approach depends on the citation network, meaning that uses of foundation models outside formal publications and those not explicitly cited are not captured, a challenge previously noted for research software~\cite{smith2016software}. Similarly, since we use the text around a citation to classify whether it is being cited/used/customized, if a model is cited once in a general context but later used without further citation, our analysis would not detect that usage. Because we analyze the text of publications, we are only able to analyze open-access literature. We apply appropriate weighting schemes to adjust for this and other data gathering limitations (see Section~\ref{sec:weighting}). 

\section{Discussion}

The adoption of AI in science is anticipated to be an enormous paradigm shift~\cite{xu2023ai4science}. In this article, we explore one of the key pathways for such a change: the adoption of foundation models. By 2024, we find that \fmUsingShareLaterYear{} of all scientific publications use foundation models without modification, while \fmCustomizingShareLaterYear{} customize them. To put these percentages in context, it is estimated that 36\% of biomedical studies use microscopy~\cite{reigoto2021comparative}, and 8–11\% of computational science papers rely on supercomputing resources~\cite{bianchini2025aisupercomputingpoweringwave}.
This suggests that foundation models could still have significant room for further adoption if they are as important as these other tools. Furthermore, the near-exponential growth trajectory in the usage of foundation models signals that this shift is still very much ongoing.


Three methodological advances distinguish our approach and enable novel insights. First, we track individual foundation models rather than AI as a monolithic category, making it possible to analyze model characteristics including size, openness, modality, and release date. We find that older, small models are predominantly adopted, and that vision models still represent the most-used type. Overwhelmingly, scientists are currently using open models, but the use of restricted-use LLMs is growing rapidly.  Second, we decompose engagement into distinct modes: citation without use, off-the-shelf usage, and active customization. This taxonomy reveals that a majority of papers discussing foundation models do not deploy them, revealing the gap between awareness and adoption. We also find remarkable differences in adoption, for example, in Linguistics, more than a third of all papers adopt foundation models. We also find clear evidence that, despite being easier to use, off-the-shelf implementations of foundation models are not yet sufficient for many scientists and thus a large minority are themselves customizing models to their particular needs. Third, our classification leverages LLMs for analysis of specific citation context rather than keyword matching or abstract classification, providing higher precision in distinguishing types of computational use from tangential references. Collectively, these advances answer fundamentally different questions: not just whether AI is being adopted, but which specific models are being used, how they are being deployed, and what characteristics make certain models more attractive for scientific applications.

While our research clearly indicates the importance of AI foundation models in science, it also highlights the yet-unrealized potential that remains. Such potential broadly falls into three categories: using more, using newer and using larger. \textit{Using more} is clear, with only 1\% of publications using foundation models, there is still enormous potential to use either general-purpose or field-specific tools more in science.  \textit{Using newer} reflects the fact that many scientists are using models 3-6 years old.  Given the enormous progress in AI performance each year~\cite{epochai2024trends}, this suggests there are significant untapped capabilities not yet being harnessed by scientists.  \textit{Using larger} captures that scientists are disproportionately using small models, despite clear evidence that large models yield better performance~\cite{rosenfeld2019constructive, kaplan2020scaling} and our own suggestive evidence that using larger models achieves better journal and citation outcomes. We look forward to the scientific discoveries that will be unlocked as this unrealized potential turns into insights about the world.



\clearpage 

\begin{ack}

The authors would like to express their sincere gratitude to Emanuele Del Sozzo for his generous support, insightful advice, and invaluable expertise in systems and hardware throughout the course of our experiments. The authors thank Zachary Brown for proofreading the paper and providing insightful comments. A.T. is grateful to Mercè Crosas for useful conversations during her stay at the Barcelona Supercomputing Center. We thank the undergraduate students who assisted with data collection for their contributions: Evan Zhang, Selinna Lin, Denis Siminiuc, Grace Yuan, Emma Li, Sri Saraf, Raul D Campos, Yibo Cheng, Alvin Banh, Dora M. Zhou, Ingrid Tomovski, Kristina Sakayeva and Maeve Zimmer. 

\subsection*{Funding}

This work is funded by Microsoft, Open Philanthropy/Good Ventures and the Alfred P. Sloan Foundation (G-2025-25164). We acknowledge support from OpenAI Research Credits \& the UROP Program at MIT. 
The authors acknowledge the MIT SuperCloud and Lincoln Laboratory Supercomputing Center for providing HPC resources that have contributed to the research results reported within this paper.

\subsection*{Authors contributions}

A.T. conceptualized the study and acquired funding, collected the initial dataset, generated the visualizations, performed preliminary analysis, and wrote the initial and subsequent drafts of the manuscript.
A.F. developed the data pipeline software, collected and curated the dataset, performed formal data analysis, generated the visualizations and text statistics, and contributed to writing the manuscript.
J.S. conducted data collection and integration, contributed to the preliminary analysis, and supported funding acquisition efforts.
N.T. conceptualized the study and acquired funding, supervised the project and the methodological choices, provided guidance on data collection and visualization, and contributed to the writing of the manuscript.

\subsection*{Competing interests}

The authors declare no competing interests.

\end{ack}





\medskip

{
\small
\bibliographystyle{unsrtnat}
\bibliography{bibliography}
}


\newpage


\appendix

\renewcommand{\thefigure}{A\arabic{figure}}
\renewcommand{\thetable}{A\arabic{table}}
\renewcommand{\theequation}{A\arabic{equation}}
\setcounter{figure}{0}
\setcounter{table}{0}
\setcounter{equation}{0}


\section{Data and materials availability}

The dataset and executable notebooks underlying this analysis are part of an ongoing research program and will be released publicly in full once the remaining studies drawing on them have been published.


\section{Materials and Methods}

\renewcommand{\P}{\mathbb{P}}

\subsection{Data Sources}
\label{sec:datasources}

Our analysis aggregates and extracts data from multiple sources. We begin by curating a custom dataset of foundation models and enhancing it with relevant metadata. We then retrieved papers which cite these foundation models, enabling us to extract key citation sentences and paper affiliations. 

\subsubsection{Foundation Models}

The Epoch AI Index Dataset gathers a collection of foundational machine learning models and systems, many of which formally recognized as AI milestones~\cite{EpochNotableModels2024}. The model collection was guided by specific criteria, including the presence of a machine learning element, demonstration of experimental findings, and indication of a contribution to advance the field of AI~\cite{sevilla2022compute}. The majority of the models meet at least one criterion of significance, like receiving over 1000 citations, possessing historical relevance, contributing to a notable advancement in the field, being employed in an important setting, or fulfilling certain subjective criteria for newer models. The dataset is assembled from a variety of sources, such as highly cited papers from leading AI conferences, existing related datasets, literature reviews, Papers with Code\footnote{https://paperswithcode.com}, and recommendations from experts~\cite{sevilla2022compute}. 

We obtained a snapshot of the Epoch AI Index dataset, covering \numfoundationmodels{} models released between 1962 and 2024, including notable foundation models like LLaMA~\cite{touvron2023llama}, GPT-4~\cite{achiam2023gpt}, and Claude~\footnote{https://claude.ai}. Given the broad scope of the Epoch AI Index dataset, which includes foundation models, optimization algorithms, and model architectures, we applied a custom definition (see Section \textit{Foundation Model Definition}) to identify and extract only those foundation models relevant to this study. This definition was deliberately designed to include historically significant releases such as AlexNet~\cite{krizhevsky2012imagenet}, while excluding purely architectural or theoretical contributions, such as the original Transformer~\cite{vaswani_attention_2023}, that were neither released nor verifiably used as pre-trained models.

Further, the original snapshot of the Epoch AI Index dataset lacked information on model size, model openness, and other relevant factors for many models. To improve its completeness, undergraduate students from \ourinstitution{} assisted in reviewing model entries, classifying the models, and adding missing information where possible. Model size was defined as the total number of trainable parameters. Model availability is classified based on whether the model can be downloaded, accessed via an API, or is entirely unavailable. We classify models as \textit{open foundation models} (open weight models) if weights are publicly downloadable and modifiable, \textit{restricted-access} if they are available only via API, and \textit{unreleased} if they are not accessible for reuse. Our dataset also contains undocumented models for which openness could not be determined.

Many of these models were released alongside a paper publication, while others were introduced through blog posts or software repositories. For models documented with a paper URL, we located their Semantic Scholar Corpus IDs in the Semantic Scholar Academic Graph (S2AG) database to retrieve publication metadata. This metadata included details such as the paper's title, abstract, citation count, and a list of citing papers, facilitating the next step of our analysis, which was retrieving the citing papers.

\subsubsection{Citing Papers} \label{sec:s2ag}

We leverage the S2AG: an extensive, open-access database that encompasses a vast collection of scientific publications and related academic information. It serves as a rich index of scientific literature, offering detailed metadata on hundreds of millions of academic articles and books. S2AG provides data about papers, authors, citations, and venues on over 200 million research papers across all fields of study~\cite{kinney2023semantic, wade2022semantic}. It is intended as a free and open resource for the research community, offering both a downloadable dataset and access through APIs. When constructing our dataset, we utilize the citation graph, paper metadata (i.e., title, publication year, field of study), pre-extracted paper plain-text, and open-access download links. We retrieve and work with the S2ORC corpus from March 12th, 2025.


Semantic Scholar, along with the S2AG dataset, provides plain-text formatted version of millions of open-source papers through the S2ORC dataset~\cite{lo-etal-2020-s2orc, lo2019s2orc} enhanced with citation annotations and download links to the open-access PDF files. We aim to maximally expand our coverage of the academic literature: we do this by capturing both the pre-extracted plain text from S2ORC, while also downloading and processing open-access PDFs through our own custom pipeline. Where available for download, we parse the PDF files using Meta's Nougat model~\cite{blecher2023nougat} which has the distinct advantage over other models we tested (e.g. Marker\footnote{https://github.com/VikParuchuri/marker}) of being trained on specifically academic papers and standardizing bibliographies. Our text conversion is followed by extraction of in-text citations and affiliations sections using regular expressions. When available, our analysis defaults to the S2ORC plain-text, which includes annotations for these features. The text for in-text citations and affiliations are processed later in our methodology to create a structured dataset.

We find \BothSSUSProportion{} of papers in our final dataset are detected in both pipelines, \SSProportion{} is detected by just S2ORC plain-text, and \USProportion{} detected by our custom built pipeline. Our final sample covers \numuniquepapers{} unique papers citing foundation models through paper citations. Across a randomly selected set of 50 citation sentences per pipeline, the false positive rate for both methods was 0\%, giving high confidence in the fidelity of our detection mechanism.




Our initial analysis using the S2ORC API to retrieve references indicated that 62\% of open-access papers lacked this annotated data, despite the presence of complete metadata. This suggested a significant issue with the completeness of the S2ORC plain text papers, leading us to develop our own custom pipeline. After developing and processing our own extracted in-text citations, it was discovered that these missing references resulted from an API bug, where even modest batch sizes (200 and above) sporadically returned missing results, while smaller batch sizes did not. The bug has since been reported to the Semantic Scholar team. 


\subsection{Methods}
\label{matmethods}
To understand how each foundation model is referenced within the citing papers and the scientific literature more generally, we apply state of the art large language models to classify the citing sentences, structure affiliations metadata, and categorize institutions. Moreover, to adjust for both closed-access publications and open-access publications which are not captured through our pipeline, we employ sample up-weighting when presenting our results.


\subsubsection{Field Selection}
\label{ref:fieldselection}

When selecting fields to present in our tables and graphics, we apply a uniform criteria to ensure the reliability of our results. In particular, we only include fields if their minimum sample weight in 2024 is less than 15; that is, we exclude results where each 2024 sample within a field requires $15\times$ or more upsampling (see section \ref{sec:weighting} below for our weighting scheme). Fields with higher sample weights yield results which are less robust to individual sample variation. Those fields are \omittedFields{}. For completeness, Figure~\ref{fig:overall_fp} shows adoption over time in these fields.

\subsubsection{Citation Classification} 
Citation context analysis is an important task in natural language processing that examines the ways and reasons scholars reference each other's work. This task is framed as a multi-class classification problem~\cite{jurgens2018measuring}, where each class signifies a distinct citation intent. Early research focused on single-sentence citation contexts, but it was later recognized that a broader context window is essential for a more accurate understanding~\cite{athar2012context, lauscher2021multicite}. Recent methods employ a citation labeling scheme that categorizes citations into classes such as \textit{background}, \textit{motivation}, \textit{similarities}, \textit{differences}, \textit{uses}, and \textit{extends}~\cite{hu2022varmae, cohan2019structural, lauscher2021multicite, lahiri2023citeprompt}.

\begin{figure}
    \centering
    \includegraphics[width=0.5\linewidth]{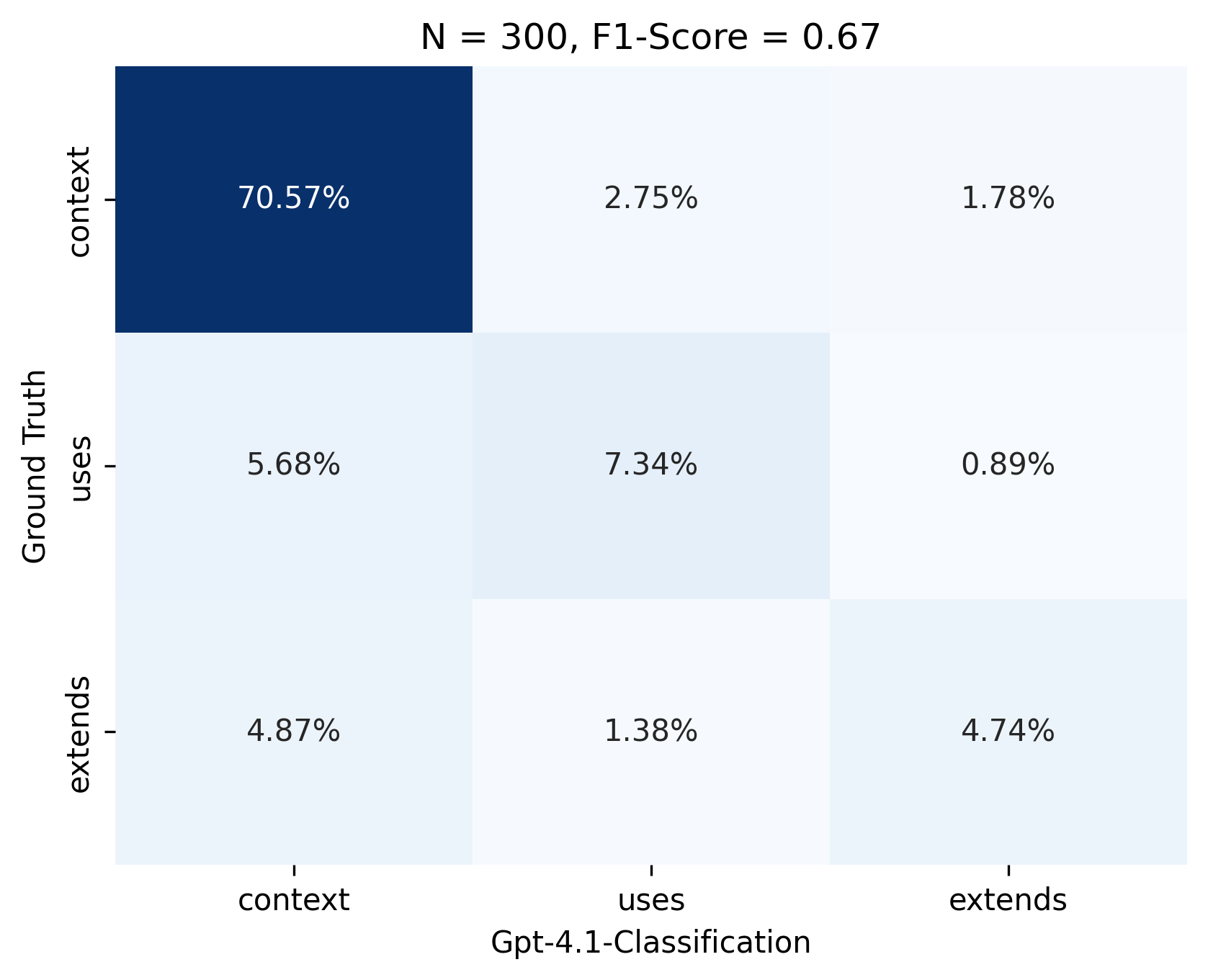}
    \caption{gpt-4.1-mini classifier confusion matrix, column-weighted by observed distribution.}
    \label{fig:cm}
\end{figure}

Following the mixed success of existing methods like \texttt{MultiCite}~\cite{lauscher2021multicite} for our problem, we turned to language models and focused on three classification categories inline with our research question: \textit{background},  \textit{uses}, and \textit{extends}. This classification allows us to distinguish how researchers reference models, whether citing foundational work, applying models, or customizing them for specific needs. We experimented extensively with various classification methods, including existing small language models like MultiCite, position-based classification (e.g. all introduction citations are ``background"), open-source language models, large proprietary language models, chain-of-thought and question-answer sequences, and training small feed-forward networks on the output of the aforementioned methods. In addition, we conducted citation-intent classification using full-paper context with OpenAI’s GPT-4 models, chosen for their cost-efficiency and context window capabilities at the time of the experiment. Among all approaches, we found that a three-sentence context classified using zero-shot prompting of OpenAI’s GPT-4.1-mini~\cite{OpenAIGPT4omini2024} delivered the best results. The resulting confusion matrix (Figure~\ref{fig:cm}), was produced by labeling 100 randomly selected sentences for each classification category and weighting the columns according to our dataset’s empirical distribution. We used the following prompt:\\


\noindent\textbf{System Instruction}
\begin{quote}
You are an expert in many areas of the scientific literature, with a specialty in how machine learning models are used.
\end{quote}

\vspace{1em}

\noindent\textbf{Prompt Template}
\begin{quote}
You will pretend to be the author of some sentences from an academic paper which reference \{model\_descriptor\}. 
Your goal is to determine the extent to which you used the cited model in your paper. 
The specific citation which references the foundation model is highlighted using HTML style $\langle$cite$\rangle$ brackets. 

Choose from the following list to determine the extent to which you adopted the model:

(1) I merely referenced the cited model as relevant background, for its methodology, or its dataset.\\
(2) I use the cited model or a part of the model itself, but I don't alter the weights.\\
(3) I make updates to the cited model's weights, through additional gradient based training such as fine-tuning.\\

Please respond in JSON format only, with key ``answer`` and value either 1, 2, or 3.

The sentences are as follows:

``\{multisentence\}``
\end{quote}


\subsubsection{Model Version Disambiguation}
\label{sec:disambiguation}

For many foundation model papers, multiple models are introduced within a single paper. When citing these papers, the citation graph alone leaves ambiguity as to which specific model is being referenced. To mitigate this, we group our foundation model list by their Semantic Scholar Corpus ID, noting when a single ID has multiple associated model keys. For those such citations, we disambiguate the individual model keys for each citation using Llama3.1:8B. The language model is prompted using (a) the name of the class of models (e.g. ``Llama"), (b) the list of specific model keys (e.g. ``llama\_70B", ``llama\_7B"), and (c) the full three-sentence context of the citation. The model is allowed to use the key ``UNCLEAR" if the context is still ambiguous. In our final analysis, unclear sentences split the associated weight per sample to each model key according to the distribution of unambiguous sentences (or evenly, in the rare case of all samples being ambiguous). 

\subsubsection{Extracting Paper Affiliations} \label{sec:affiliations} To obtain accurate information on adopters’ affiliations and countries, we extracted the first section of each paper listing the authors and their institutions. We employ Meta's Llama 3.1 8B open-source language model~\cite{dubey2024llama, grattafiori2024llama} to (a) return the content of these sections in a normalized, structured JSON format and (b) to infer the country of origin for each institution, as well classifying the organization as either academic or industry. We validate the structure of these outputs using Pydantic~\cite{pydantic}. 

\subsubsection{Hardware and Computational Resources}
Across our pipeline, we have various uses for GPU acceleration: conversions of PDFs to Markdown using Nougat (described in Section~\ref{sec:s2ag}), extraction of structured paper affiliations using \textsc{Llama 3.1} (Section~\ref{sec:affiliations}), disambiguation of model versions within citing sentences (Section~\ref{sec:disambiguation}), and initial experiments with various citation classification techniques. For these computations, we used NVIDIA V100 Tensor Core GPUs and NVIDIA A100 Tensor Core GPUs distributed across 32 GPU Nodes on \ourhpc. 

\subsubsection{Paper-Level Classification and False Positive Corrections}
\label{sec:falsepositives}

When we analyze citation behavior, we classify individual citing sentences as reflecting one of three types of model engagement: \textsc{context} for mere background citations, \textsc{uses} for deploying models unchanged, and \textsc{extends} for model customization. When aggregating sentences within paper, we assign the paper a label if at least one of its citation sentences matches the category: a paper is labeled \textsc{extends} if any sentence is classified as such; \textsc{uses} if no extends are found but at least one \textsc{uses} is present; and \textsc{context} otherwise. Adoption is defined as either extends or uses.

A challenge arises as papers with more citation sentences are more likely to be (mis)classified into one of these categories, as each additional sentence increases the chance of a \emph{false positive}. In contrast, estimating false positives at the sentence level is straightforward: we can take a random sample and manually check the accuracy. At the paper level, however, aggregation introduces bias that must be modeled.

To correct for this, we use a Bayesian approach. Let $FP(p)$ denote the event that paper $p$ is a false positive. Let $\hat{e}$ and $\hat{u}$ be the observed counts of extends and uses sentences, respectively, and let $\P_n$ denote probabilities conditioned on a paper containing $n$ citing sentences. Then by Bayes’ theorem:

\begin{align*} 
\underbrace{\P_n(FP(p))}_{\textbf{desired}} \cdot 
\underbrace{\P_n(\hat{e} = m,\hat{u} = k \mid FP(p))}_{\text{modeled via Multinomial}} 
= 
\underbrace{\P_n(FP(p) \mid \hat{e} = m,\hat{u} = k)}_{\text{sampled empirically}} 
\cdot 
\underbrace{\P_n(\hat{e} = m, \hat{u} = k)}_{\text{observed directly}} .
\end{align*}

By estimating $\P_n(FP(p))$ for different kinds of paper-level false positives, we can down-weight the total paper counts when grouping by classifications. We consider adopts and extends false positives as two separate cases (from which uses false positives can be derived), and outline below how the conditional terms above are approximated to derive our desired false positive rate, $P_n(FP(p))$, for each false positive type. Finally, we explain when and how these adjustments are incorporated into our results.

\paragraph{Binomial Model Given Paper False Positives}

\begin{figure*}[t!]
    \centering
    \includegraphics[width=\linewidth]{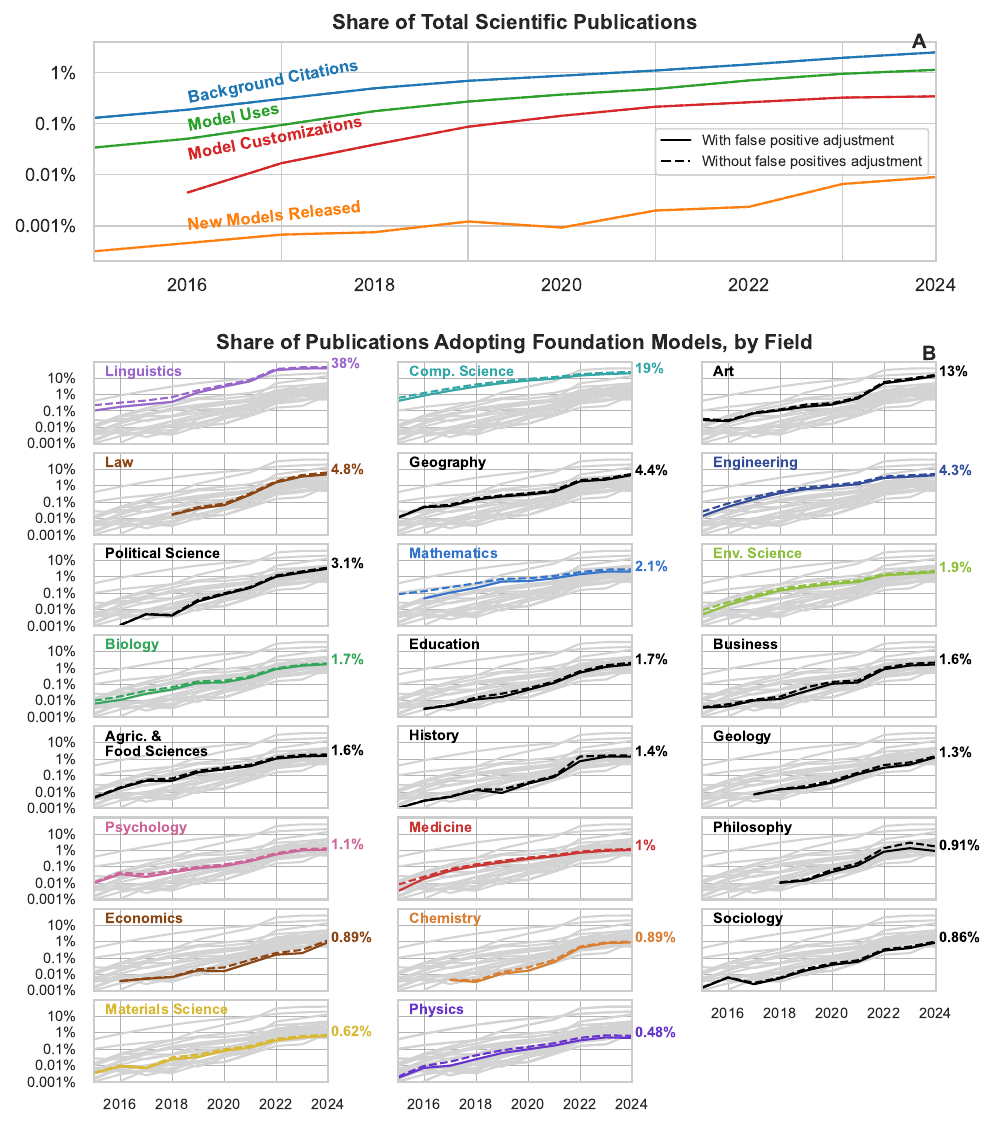}
    \caption{Figure 1 with false positive adjusted and unadjusted proportions. Includes fields which did not meet our selection criterion, but are included for completeness. Solid lines are adjusted, dashed lines are unadjusted.}
    \label{fig:overall_fp}
\end{figure*}

We first aim to approximate the distribution of sentences labeled as uses/extends within a paper of size $n$, conditioned on this paper being a false positive. The parameters of the multinomial distribution are derived from hand labeled sentence classifications samples, using values which we argue result in upper bounds on the false positive rates. By using upper-bound false positive rates, we systematically err on the side of conservatism in our aggregated counts—that is, our corrections are designed to underestimate rather than overestimate the true prevalence. For transparency, we also include uncorrected counts in Figure~\ref{fig:overall_fp}.


\subparagraph{Adoption false positives} are papers which contain no true uses or extends sentences, yet have at least one (possibly more) false positives for either class. We denote the statement ``paper $p$ is an adoption false positive" as $FP_a(p)$. We use the \textit{context to uses} and \textit{context to extends} false positive rates from our sentence level confusion matrix ($r_{c \rightarrow u}$ = 2.8\%, $r_{c \rightarrow e}$ = 1.8\%) as parameters in a 3-class multinomial distribution:

\[\P_n(\hat{u} = k, \hat{e} = m | FP_a(p)) = \frac{Multinom(n, r_{c \rightarrow u}, r_{c \rightarrow e})(k,m)}{1 - Multinom(n, r_{c \rightarrow u}, r_{c \rightarrow e})(0,0)}\]

Since our estimates only involve cases where $k + m = 1$, we can (and do) use an equivalent binomial with rate $(r_{c \rightarrow u} + r_{c \rightarrow e})$ to estimate $\P_n(\hat{u} + \hat{e} = 1 | FP_a(p))$.

\subparagraph{Extends false positives ($FP_e$)} are papers with no true extends sentences and at least one false positive extends label. We denote the statement ``paper $p$ is an extends false positive" as $FP_e(p)$. Note that these papers may be true usage papers, and we therefore must consider both \textit{context-to-extends} and \textit{uses-to-extends} false positive rates. Unlike \textit{adopts} false positives, our confusion matrix does not provide enough information to estimate the Bernoulli probability of the binomial, since the true distribution contains not only true \textit{context} sentences, but true \textit{uses} sentences as well, the distribution of which is unknown. Therefore, transition probabilities are insufficient to glean a precise estimate.

The true parameter is necessarily between $r_{c \rightarrow e}$ and $r_{u \rightarrow e}$. In order to \textit{upper bound} our false positive rate, we want to \textit{lower bound} this binomial term. As we discuss below, our estimation uses $m = 1$ (or $\P_n(\hat{e} = 1 | FP_e(p))$), and therefore \textit{upper bounding} the binomial rate will minimize this value (which can be thought of as minimizing the likelihood of getting only one false positive). We thereby use a probability of $r_{c \rightarrow e} + r_{u \rightarrow e}$ for our binomial distribution, normalized since $\hat{e} \neq 0$ when conditioning on a false positive:

\[\P_n(\hat{e} = m | FP_e(p)) = \frac{Binom(n, r_{c \rightarrow e} + r_{u \rightarrow e})(m)}{1 - Binom(n, r_{c \rightarrow e} + r_{u \rightarrow e})(0)}\]



\paragraph{Sampling for Conditional False Positive Rate}

To estimate the term $\P_n(FP(p) \mid \hat{e}=m, \hat{u}=k)$, we select papers with specific $(m,k)$ configurations and manually label their sentences to obtain an empirical estimate of the false positive probability. In particular, we leverage the most sample efficient $m,k$ characteristics, using $(m,k) \in \{(0,1),(1,0)\}$ for adoption false positives, and $m = 1$ for extends, where $k$ varies randomly. This allows us to determine the false positive rate with only one label per paper, and therefore gives us a more robust sample size per unit of labeling time.

More precisely, for $n \in \{1, 3, 5, 9, 13\}$, we label $20$ sentences for each of the three characteristics: $(m,k) = (1,0)$, $(m,k) = (0,1)$, and $m = 1$. For extends false positives, we observe directly from our sample $\P_n(FP(p) \mid \hat{e}=1)$, whereas for adoption false positives, we directly observe values $\P_n(FP(p) \mid \hat{e}=1, \hat{u}=0)$ and $\P_n(FP(p) \mid \hat{e}=0, \hat{u}=1)$, weighting them by their frequencies in the dataset to obtain the more general $\P_n(FP(p) \mid \hat{e} + \hat{u} = 1)$.

The resulting probabilities, each conditioned on some number of sentences $n$, are linearly extrapolated point-by-point to fill in missing $n$ values. At this point, we have an empirical estimation of $\P_n(FP(p))$, for both adopts and extends false positives. We now explain how these are incorporated into our results.

\paragraph{Adjustments to Aggregated Counts}

When reporting \textit{total counts} of papers which use, extend, or generally adopt models, we adjust our total counts based on the estimated false positive rates. Consider some subset of papers, for example, Computer Science papers from 2020. If we call $f(n)$ the proportion of papers in this subset which have $n$ sentences, then we can compute the overall false positive rate as follows:

$$\P(FP(p)) = \sum f(i) \cdot \P_i(FP(p)) $$

Thus given a subset of paper containing all classifications with total sample weight $W$, and total classification weights $W = W_c + W_u + W_e$, we make the following adjustments: 

\begin{align*}
    W_c &\rightarrow W_c + W \cdot \P(FP_a(p)) \\
    W_u &\rightarrow W_u + W \cdot \P(FP_e(p)) - W \cdot \P(FP_a(p)) \\
    W_e &\rightarrow W_e - W \cdot \P(FP_e(p))
\end{align*}

Similarly, when aggregating extends and uses into one adoption category, our transformation is equivalent to summing $W_u$ and $W_e$ in the transformations above.

In Figure \ref{fig:overall_fp} we display Figures \ref{fig:overall} before and after making false positive adjustments.

\paragraph{Using Unadjusted Counts for Growth Rates}

For data subsets with particularly low rate of adoption, usage, or customization, on rare occasions we find our expected false positive correction is greater than the total number of observations. We avoid reporting absolute figures in these cases, but report qualified growth rates (Table~\ref{tab:fields}). 

\paragraph{False Negatives}

Our corrections are limited to false positives, and do not include false negatives. To see why, consider two potential effects: a false negative rate of 5\% and false positive rate of 5\%.  By 2024, the share of publications that customize foundation models is \fmCustomizingShareLaterYear{}. If there is a false negative rate, then this number should actually be roughly 0.32\%, which does not meaningfully change our answers. By contrast, publications that cite foundation models only as background (\fmCitingShareLaterYear{}) or that use them without modification (\fmUsingShareLaterYear{}) together form a pool roughly twelve times as large, and a false positive in either category is counted as a customization. A 5\% false positive rate applied to that pool would therefore make the underlying customization rate appear to be roughly 0.48\%, or 60\% larger than it should be. 

More specifically, the average number of sentences in adoption papers is \AverageSentencesAdoptPapers{}, with \AverageAdoptsSentencesAdoptPapers{} of those sentences exhibiting adoption on average. This introduces an asymmetry in our favor, in which all adoption sentences must be false negatives in order for a paper to be a false negative, whereas with false positives, only one such false positive sentence is needed. Also, when aggregating averages within \textsc{uses} or \textsc{extends} classification, false negatives do not affect our outcomes provided they are not systematic for our desired metric (e.g. model size, model openness), which is untrue of false positives. The reason is that a false negative merely removes a paper that genuinely belonged, leaving what remains a clean if slightly smaller sample, whereas a false positive admits a paper whose model size and openness were never those of an adopter at all. Finally, in our manually labeled sample, false negatives were generally more ambiguous to the labeler than false positives.

\subsubsection{Weighting Schema}
\label{sec:weighting}

Our primary dataset consists of open-access academic papers. To account for the broader landscape of scientific literature, we leverage the full citation graph from Semantic Scholar and apply a weighting schema that allows us to adjust our estimates to the full population of publications.

Fix a year $T$ and a foundation model $M$. There are two ways of determining the number of unique papers citing $M$ which are published in year $T$: (1) through the S2AG dataset, which includes the widest swath of academic papers, which we denote $Y^M_T$; and (2) observationally through our sample, which is limited to open-access publications extracted through either of our extraction pipelines, which we denote $X^M_T$. Thus for each sample (that is, each sentence from a paper published in year $T$ citing model $M$), we could assign a weight given by:
$$W_T^M = \frac{Y^M_T}{X^M_T}$$
However, a small fraction of papers are missing publication years. For each set of foundation model citers, we adjust our weights such that the distribution of years among papers without year metadata matches the distribution among papers where we do have year metadata. Let $T = 0$ index a year which is missing, and let $p_T$ be the proportion of papers citing $M$ with known years from year $T$ (that is for $T > 0$, $p_T = X_T^M/\sum_{\tau \neq 0}  X^M_\tau$). Then the property we desire from our weights is that for all $t \neq 0$:
$$W_t^M\left[X_t^M + p_tX_0^M\right] = Y_t^M$$

Intuitively, this means that the weight from our known sample, together with a proportional share of the unknown sample, should sum to the correct total number of papers. To satisfy this constraint, we augment the naive weight and also define a weight for papers with missing years:

\begin{equation*}
    W_t^M = \begin{cases}
\displaystyle \frac{Y^M_t}{X_t^M + p_tX_0^M} & t \neq 0 \\
\\
\displaystyle \sum_{\tau \neq 0} p_\tau W_\tau^M & t = 0
\end{cases}
\end{equation*}

Finally, we note that the estimation of $Y_t^M$ implicitly relies on an additional assumption, since the S2AG may contain missing years as well. To address this, we apply a smoothing procedure to account for missing years in  $Y_t^M$ as well, assuming they follow the same distribution, thus $Y_t^M = Z_t^M + q_tZ_0^M$, where $Z$ is the true distribution of across years, $Z_0^M$ is the number of $M$ citers missing a publication year, and $q_t$ is defined similarly to $p_t$.

Empirically, we find that weights within a paper are generally quite similar: the standard deviation of weights within a paper's citing sentences is, on average, less than $2$. Among different foundation models cited by the same paper, we don't expect the availability of papers citing those models to vary substantially, thus this is consistent with expectations. When aggregating to the paper level -- that is, aggregating information from multiple classified sentences to one paper -- we always use the maximum weight among all the paper's sentences for that sample.  Perhaps counterintuitively, this is a \textit{lower bound} on the true number of papers represented by this sample. Consider $\mathcal{M}$ to be the set of models cited by some paper $p$. For a particular sentence citing model $m \in \mathcal{M}$, the weight tell us how many unseen papers also citing $m$ are represented by the paper $p$. However, if the paper cites two (or more) distinct models $m_1, m_2 \in \mathcal{M}$, then $p$ represents the unseen papers citing $m_1$ (denoted $\mathcal{P}_{m_1}$) and the unseen papers citing $m_2$ (denoted $\mathcal{P}_{m_2}$). By the union bound, we clearly have $\max(|\mathcal{P}_{m_1}|, |\mathcal{P}_{m_2}|) \leq | \mathcal{P}_{m_1} \cup \mathcal{P}_{m_2}| \leq |\mathcal{P}_{m_1}| + |\mathcal{P}_{m_2}|$.





\section{Supplementary Tables}

\begin{table}[tbp]
    \centering
    \resizebox{\textwidth}{!}{%
\begin{tabular}{cccccccccc}
\toprule& \multicolumn{6}{c}{\textbf{Model Adoption}} & \multicolumn{1}{c}{\textbf{Model Usage}} & \multicolumn{2}{c}{\textbf{Model Customization}} \\ \cmidrule(lr){2-7} \cmidrule(lr){8-8} \cmidrule(lr){9-10}

\textbf{\makecell[t]{Field of\\Study}} & \textbf{\makecell[t]{Model Parameters\\ Avg (Q1, Q3)}} & \textbf{\makecell[t]{Top 3\\Foundation\\Models}} & \textbf{\makecell[t]{Model Age in Years\\Avg (Q1, Q3)}} & \textbf{\makecell[t]{Total Unique\\Models Adopted}} & \textbf{\makecell[t]{Open-Weight}} & \textbf{\makecell[t]{Adoption\\by 2024}} & \textbf{\makecell[t]{Usage Growth\\(3-yr CAGR)}} & \textbf{\makecell[t]{Customizations Growth\\(3-yr CAGR)}} & \textbf{\makecell[t]{Customizers\\(\% of Adopters)}} \\ \hline\midrule

\makecell[t]{Computer Science} & \makecell[t]{137M\\(12M,2B)} & \makecell[t]{ResNet-152 (ImageNet) (12\%)\\BERT-Large (8\%)\\VGG16 (6\%)\\} & \makecell[t]{4.7\\(3.0, 6.0)} & \makecell[t]{857} & \makecell[t]{77\%} & \makecell[t]{\ComputerScienceAdoptsCurrentRate{}} & \makecell[t]{\ComputerScienceUseCagr{}\ComputerScienceUseExtendUnadjusted{}} & \makecell[t]{\ComputerScienceExtendsCagr{}\ComputerScienceUseExtendUnadjusted{}} & \makecell[t]{41\%} \\ \hline

\makecell[t]{Linguistics} & \makecell[t]{635M\\(55M,7B)} & \makecell[t]{BERT-Large (26\%)\\xlm-roberta (6\%)\\RoBERTa-Large (6\%)\\} & \makecell[t]{3.3\\(2.0, 4.0)} & \makecell[t]{357} & \makecell[t]{86\%} & \makecell[t]{\LinguisticsAdoptsCurrentRate{}} & \makecell[t]{\LinguisticsUseCagr{}\LinguisticsUseExtendUnadjusted{}} & \makecell[t]{\LinguisticsExtendsCagr{}\LinguisticsUseExtendUnadjusted{}} & \makecell[t]{47\%} \\ \hline

\makecell[t]{Engineering} & \makecell[t]{57M\\(7M,452M)} & \makecell[t]{ResNet-152 (ImageNet) (14\%)\\VGG16 (7\%)\\faster\_r-cnn (4\%)\\} & \makecell[t]{5.5\\(4.0, 7.0)} & \makecell[t]{440} & \makecell[t]{76\%} & \makecell[t]{\EngineeringAdoptsCurrentRate{}} & \makecell[t]{\EngineeringUseCagr{}\EngineeringUseExtendUnadjusted{}} & \makecell[t]{\EngineeringExtendsCagr{}\EngineeringUseExtendUnadjusted{}} & \makecell[t]{40\%} \\ \hline

\makecell[t]{Mathematics} & \makecell[t]{103M\\(8M,1B)} & \makecell[t]{ResNet-152 (ImageNet) (19\%)\\VGG16 (8\%)\\ResNet-110 (cifar-10) (6\%)\\} & \makecell[t]{5.2\\(3.0, 7.0)} & \makecell[t]{280} & \makecell[t]{68\%} & \makecell[t]{\MathematicsAdoptsCurrentRate{}} & \makecell[t]{\MathematicsUseCagr{}\MathematicsUseExtendUnadjusted{}} & \makecell[t]{\MathematicsExtendsCagr{}\MathematicsUseExtendUnadjusted{}} & \makecell[t]{46\%} \\ \hline

\makecell[t]{Environmental Science} & \makecell[t]{56M\\(10M,302M)} & \makecell[t]{ResNet-152 (ImageNet) (13\%)\\VGG16 (7\%)\\AlphaFold (2) (6\%)\\} & \makecell[t]{5.4\\(4.0, 7.0)} & \makecell[t]{329} & \makecell[t]{81\%} & \makecell[t]{\EnvironmentalScienceAdoptsCurrentRate{}} & \makecell[t]{\EnvironmentalScienceUseCagr{}\EnvironmentalScienceUseExtendUnadjusted{}} & \makecell[t]{\EnvironmentalScienceExtendsCagr{}\EnvironmentalScienceUseExtendUnadjusted{}} & \makecell[t]{41\%} \\ \hline

\makecell[t]{Medicine} & \makecell[t]{73M\\(10M,555M)} & \makecell[t]{ResNet-152 (ImageNet) (14\%)\\VGG16 (7\%)\\densenet-264 (5\%)\\} & \makecell[t]{5.4\\(4.0, 7.0)} & \makecell[t]{410} & \makecell[t]{78\%} & \makecell[t]{\MedicineAdoptsCurrentRate{}} & \makecell[t]{\MedicineUseCagr{}\MedicineUseExtendUnadjusted{}} & \makecell[t]{\MedicineExtendsCagr{}\MedicineUseExtendUnadjusted{}} & \makecell[t]{45\%} \\ \hline

\makecell[t]{Physics} & \makecell[t]{69M\\(8M,615M)} & \makecell[t]{ResNet-152 (ImageNet) (16\%)\\VGG16 (7\%)\\image-to-image\_cgan (4\%)\\} & \makecell[t]{6.0\\(4.0, 8.0)} & \makecell[t]{182} & \makecell[t]{67\%} & \makecell[t]{\PhysicsAdoptsCurrentRate{}} & \makecell[t]{\PhysicsUseCagr{}\PhysicsUseExtendUnadjusted{}} & \makecell[t]{\PhysicsExtendsCagr{}\PhysicsUseExtendUnadjusted{}} & \makecell[t]{45\%} \\ \hline

\makecell[t]{Biology} & \makecell[t]{101M\\(17M,622M)} & \makecell[t]{AlphaFold (2) (41\%)\\ResNet-152 (ImageNet) (6\%)\\alphafold-multimer (6\%)\\} & \makecell[t]{4.0\\(2.0, 5.0)} & \makecell[t]{289} & \makecell[t]{85\%} & \makecell[t]{\BiologyAdoptsCurrentRate{}} & \makecell[t]{\BiologyUseCagr{}\BiologyUseExtendUnadjusted{}} & \makecell[t]{\BiologyExtendsCagr{}\BiologyUseExtendUnadjusted{}} & \makecell[t]{24\%} \\ \hline

\makecell[t]{Chemistry} & \makecell[t]{117M\\(18M,749M)} & \makecell[t]{AlphaFold (2) (56\%)\\alphafold-multimer (7\%)\\alphafold\_3 (3\%)\\} & \makecell[t]{3.3\\(2.0, 4.0)} & \makecell[t]{160} & \makecell[t]{89\%} & \makecell[t]{\ChemistryAdoptsCurrentRate{}} & \makecell[t]{\ChemistryUseCagr{}\ChemistryUseExtendUnadjusted{}} & \makecell[t]{\ChemistryExtendsCagr{}\ChemistryUseExtendUnadjusted{}} & \makecell[t]{13\%} \\ \hline

\makecell[t]{Psychology} & \makecell[t]{399M\\(24M,7B)} & \makecell[t]{BERT-Large (17\%)\\ResNet-152 (ImageNet) (7\%)\\RoBERTa-Large (6\%)\\} & \makecell[t]{4.5\\(2.0, 6.0)} & \makecell[t]{197} & \makecell[t]{79\%} & \makecell[t]{\PsychologyAdoptsCurrentRate{}} & \makecell[t]{\PsychologyUseCagr{}\PsychologyUseExtendUnadjusted{}} & \makecell[t]{\PsychologyExtendsCagr{}\PsychologyUseExtendUnadjusted{}} & \makecell[t]{35\%} \\ \hline

\makecell[t]{Materials Science} & \makecell[t]{56M\\(9M,347M)} & \makecell[t]{ResNet-152 (ImageNet) (13\%)\\AlphaFold (2) (11\%)\\VGG16 (8\%)\\} & \makecell[t]{5.8\\(4.0, 8.0)} & \makecell[t]{154} & \makecell[t]{79\%} & \makecell[t]{\MaterialsScienceAdoptsCurrentRate{}} & \makecell[t]{\MaterialsScienceUseCagr{}\MaterialsScienceUseExtendUnadjusted{}} & \makecell[t]{\MaterialsScienceExtendsCagr{}\MaterialsScienceUseExtendUnadjusted{}} & \makecell[t]{43\%} \\ \hline

\makecell[t]{Economics} & \makecell[t]{318M\\(12M,8B)} & \makecell[t]{BERT-Large (15\%)\\ResNet-152 (ImageNet) (8\%)\\llama\_3-70b (7\%)\\} & \makecell[t]{4.6\\(2.0, 6.0)} & \makecell[t]{76} & \makecell[t]{74\%} & \makecell[t]{\EconomicsAdoptsCurrentRate{}} & \makecell[t]{\EconomicsUseCagr{}\EconomicsUseExtendUnadjusted{}} & \makecell[t]{\EconomicsExtendsCagr{}\EconomicsUseExtendUnadjusted{}} & \makecell[t]{34\%} \\ \hline

\end{tabular}}
    \caption{Adoption of foundation models in 2019-2024 publications, by scientific field. Model age is the difference between model and publication release dates. Asterisk indicates false-positive unadjusted data (details in Section~\ref{sec:falsepositives}).}
    \label{tab:fields}
\end{table}

\begin{table}[tbp]
    \centering
    \resizebox{\textwidth}{!}{%
\begin{tabular}{ccccccc}
\toprule

\textbf{\makecell[t]{Field of\\Study}} & \textbf{\makecell[t]{Model Parameters\\ Avg (Q1, Q3)}} & \textbf{\makecell[t]{Top 3\\Foundation\\Models}} & \textbf{\makecell[t]{Model Age in Years\\Avg (Q1, Q3)}} & \textbf{\makecell[t]{Total Unique\\Models Adopted}} & \textbf{\makecell[t]{Open-Weight}} & \textbf{\makecell[t]{Customizers\\(\% of Adopters)}} \\ \hline\midrule

\makecell[t]{Computer Science} & \makecell[t]{305M\\(20M,5B)} & \makecell[t]{ResNet-152 (ImageNet) (9\%)\\BERT-Large (7\%)\\VGG16 (3\%)\\} & \makecell[t]{4.9\\(2.0, 8.0)} & \makecell[t]{750} & \makecell[t]{80\%} & \makecell[t]{40\%} \\ \hline

\makecell[t]{Linguistics} & \makecell[t]{1B\\(100M,19B)} & \makecell[t]{BERT-Large (19\%)\\xlm-roberta (7\%)\\RoBERTa-Large (5\%)\\} & \makecell[t]{3.6\\(2.0, 5.0)} & \makecell[t]{277} & \makecell[t]{85\%} & \makecell[t]{47\%} \\ \hline

\makecell[t]{Engineering} & \makecell[t]{98M\\(10M,1B)} & \makecell[t]{ResNet-152 (ImageNet) (14\%)\\VGG16 (5\%)\\faster\_r-cnn (3\%)\\} & \makecell[t]{6.1\\(4.0, 8.0)} & \makecell[t]{360} & \makecell[t]{78\%} & \makecell[t]{39\%} \\ \hline

\makecell[t]{Mathematics} & \makecell[t]{260M\\(14M,5B)} & \makecell[t]{ResNet-152 (ImageNet) (15\%)\\VGG16 (6\%)\\vit-base\_32 (5\%)\\} & \makecell[t]{5.4\\(3.0, 8.0)} & \makecell[t]{201} & \makecell[t]{72\%} & \makecell[t]{46\%} \\ \hline

\makecell[t]{Environmental Science} & \makecell[t]{81M\\(13M,504M)} & \makecell[t]{ResNet-152 (ImageNet) (12\%)\\AlphaFold (2) (10\%)\\faster\_r-cnn (5\%)\\} & \makecell[t]{6.1\\(4.0, 8.0)} & \makecell[t]{251} & \makecell[t]{82\%} & \makecell[t]{39\%} \\ \hline

\makecell[t]{Medicine} & \makecell[t]{120M\\(12M,1B)} & \makecell[t]{ResNet-152 (ImageNet) (13\%)\\AlphaFold (2) (8\%)\\densenet-264 (4\%)\\} & \makecell[t]{5.8\\(4.0, 8.0)} & \makecell[t]{346} & \makecell[t]{79\%} & \makecell[t]{44\%} \\ \hline

\makecell[t]{Physics} & \makecell[t]{103M\\(12M,878M)} & \makecell[t]{ResNet-152 (ImageNet) (15\%)\\diffusion models (lsun bedroom) (5\%)\\VGG16 (4\%)\\} & \makecell[t]{6.2\\(4.0, 8.0)} & \makecell[t]{139} & \makecell[t]{72\%} & \makecell[t]{46\%} \\ \hline

\makecell[t]{Biology} & \makecell[t]{135M\\(21M,879M)} & \makecell[t]{AlphaFold (2) (48\%)\\alphafold-multimer (7\%)\\alphafold\_3 (4\%)\\} & \makecell[t]{4.0\\(3.0, 4.0)} & \makecell[t]{227} & \makecell[t]{88\%} & \makecell[t]{19\%} \\ \hline

\makecell[t]{Chemistry} & \makecell[t]{142M\\(22M,907M)} & \makecell[t]{AlphaFold (2) (59\%)\\alphafold-multimer (8\%)\\alphafold\_3 (4\%)\\} & \makecell[t]{3.5\\(3.0, 4.0)} & \makecell[t]{121} & \makecell[t]{90\%} & \makecell[t]{11\%} \\ \hline

\makecell[t]{Psychology} & \makecell[t]{882M\\(41M,19B)} & \makecell[t]{BERT-Large (15\%)\\Llama 2 (7B) (7\%)\\ResNet-152 (ImageNet) (6\%)\\} & \makecell[t]{4.6\\(2.0, 6.0)} & \makecell[t]{152} & \makecell[t]{79\%} & \makecell[t]{35\%} \\ \hline

\makecell[t]{Materials Science} & \makecell[t]{67M\\(9M,479M)} & \makecell[t]{AlphaFold (2) (16\%)\\ResNet-152 (ImageNet) (11\%)\\VGG16 (5\%)\\} & \makecell[t]{6.2\\(4.0, 8.0)} & \makecell[t]{111} & \makecell[t]{82\%} & \makecell[t]{40\%} \\ \hline

\makecell[t]{Economics} & \makecell[t]{695M\\(23M,21B)} & \makecell[t]{llama\_3-70b (14\%)\\BERT-Large (14\%)\\Llama (7b) (10\%)\\} & \makecell[t]{4.3\\(1.0, 6.0)} & \makecell[t]{44} & \makecell[t]{81\%} & \makecell[t]{28\%} \\ \hline

\end{tabular}}
    \caption{Model adoption data aggregated by field. This table restricts the domain to more recent papers than Table \ref{tab:fields} (since 2023).}
    \label{tab:fields_recent}
\end{table}

\begin{table}[h!]
\centering
\begin{tabular}{l c}
\hline
\textbf{Model Category} & \textbf{CAGR (2021–2024)} \\
\hline
Biology Models            & \cagrBiologyModels{} \\
Video Models              & \cagrVideoModels{} \\
Robotics Models           & \cagrRoboticsModels{} \\
Speech Models             & \cagrSpeechModels{} \\
Image Generation Models   & \cagrImageGenModels{} \\
Multimodal Models         & \cagrMultimodalModels{} \\
Vision Models             & \cagrVisionModels{} \\
Language Models           & \cagrLLMs{} \\
\hline
\end{tabular}
\caption{Annual growth rates of different model categories from 2021 to 2024.} \label{supp:model_cagrs}
\end{table}

\end{document}